\documentclass[10pt,journal,twoside]{IEEEtran}

\usepackage{graphicx}

\usepackage{hyperref}
\hypersetup{
 unicode=true,
 pdfstartview={FitH},
 colorlinks=true,
 citecolor=blue,
 linkcolor=blue,
 urlcolor=blue
}

\usepackage{multirow}
\usepackage{array}
\usepackage{booktabs}
\newcolumntype{L}[1]{>{\raggedright\let\newline\\\arraybackslash\hspace{0pt}}m{#1}}
\newcolumntype{C}[1]{>{\centering\let\newline\\\arraybackslash\hspace{0pt}}m{#1}}
\newcolumntype{R}[1]{>{\raggedleft\let\newline\\\arraybackslash\hspace{0pt}}m{#1}}

\usepackage{caption}
\usepackage{subcaption}
\usepackage{enumitem}

\usepackage[dvipsnames]{xcolor}
\definecolor{asparagus}{rgb}{0.53, 0.66, 0.42}
\definecolor{brick}{rgb}{0.71, 0.2, 0.11}
\definecolor{periwinkle}{rgb}{0.55,0.55,1.0}
\definecolor{airforceblue}{rgb}{0.36, 0.54, 0.66}
\definecolor{someturquoise}{HTML}{1EC2C8}
\definecolor{Sepia}{HTML}{671800}

\usepackage{amsmath,amssymb,amsfonts}

\usepackage[mincitenames=1,maxcitenames=1,maxbibnames=1,backend=biber,isbn=false,url=false,doi=false,giveninits=true]{biblatex} %

\title{Foundation Models in Augmentative and Alternative Communication: Opportunities and Challenges}
\author{Ambra Di Paola, Serena Muraro, Roberto Marinelli, Christian Pilato
 \thanks{A. Di Paola, S. Muraro, R. Marinelli are with Fondazione Artos, Caronno Pertusella (VA), Italy.\protect\\
 C. Pilato is with the Dipartimento di Elettronica, Informazione e Bioingegneria, Politecnico di Milano, Milano (MI), Italy (contact email: christian.pilato@polimi.it). 
 }
 }

\begin{document}

\maketitle

\markboth{}{}
\thispagestyle{headings}

\begin{abstract}
Augmentative and Alternative Communication (AAC) are essential techniques that help people with communication disabilities. AAC demonstrates its transformative power by replacing spoken language with symbol sequences. However, to unlock its full potential, AAC materials must adhere to specific characteristics, placing the onus on educators to create custom-tailored materials and symbols.
This paper introduces AMBRA (Pervasive and Personalized Augmentative and Alternative Communication based on Federated Learning and Generative AI), an open platform that aims to leverage the capabilities of foundation models to tackle many AAC issues, opening new opportunities (but also challenges) for AI-enhanced AAC. 
We thus present a compelling vision--a roadmap towards a more inclusive society. By leveraging the capabilities of modern technologies, we aspire to not only transform AAC but also guide the way toward a world where communication knows no bounds. 
\end{abstract}

\section{Introduction}

\begin{figure}[!t]
\includegraphics[width=\columnwidth]{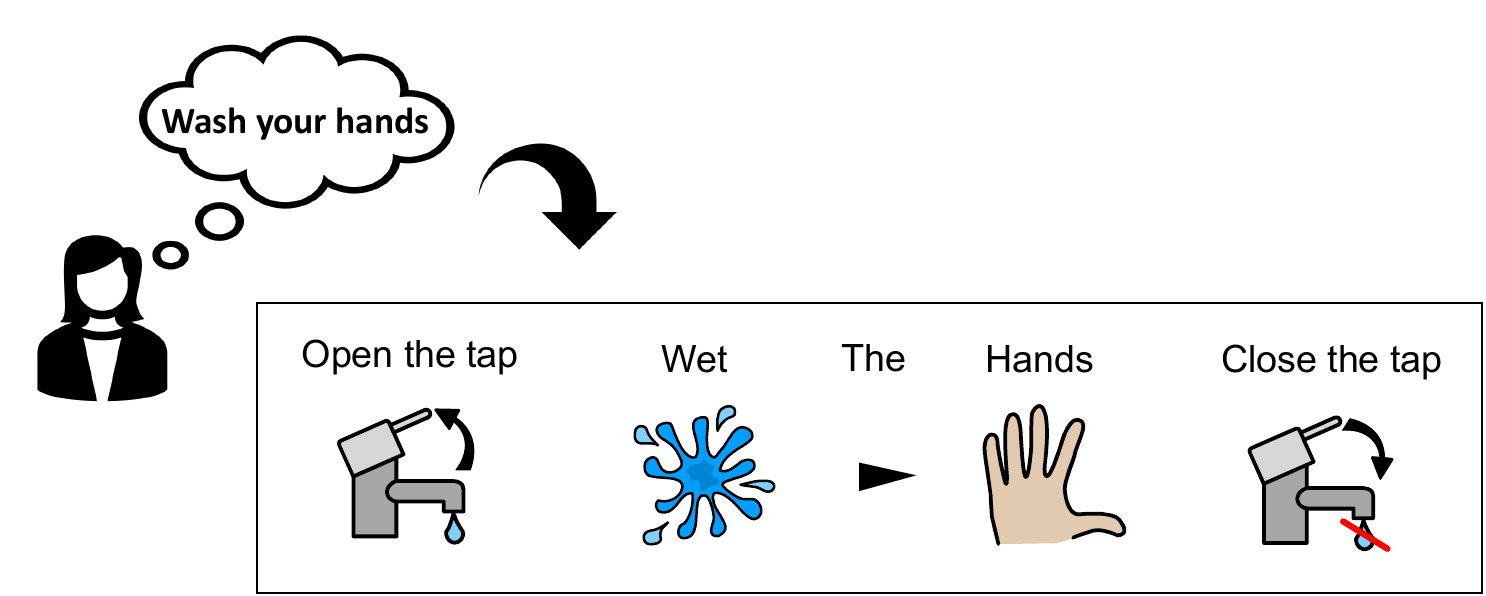}
\vspace{-12pt}\caption{Example of AAC-based communication where a sequence of symbols replaces a message. Symbols are created with Symwriter~\cite{symwriter}\vspace{-12pt}}\label{fig:aac_example}
\end{figure}

Augmentative and Alternative Communication (AAC) encompasses diverse strategies and tools meticulously designed to empower individuals facing communication disabilities, enabling them to effectively convey their thoughts, needs, and emotions~\cite{CUMMINGS202312}. These methods range from using symbol sequences, where a sequence of visual symbols or icons represents words or ideas (as shown in \autoref{fig:aac_example}), to integrating specialized assistive technology devices, such as communication boards, speech-generating devices, or eye-tracking systems.

AAC is not a one-size-fits-all solution; it requires a personalized approach. Symbols, methods, and tools should be chosen carefully to match the individual's unique needs and communication abilities. Educators, caregivers, and speech-language professionals are vital in this process. They must work together to assess, select, and implement AAC methods that suit the individual's abilities and goals~\cite{aac}.
This personalized approach goes beyond just choosing symbols or devices; it includes regularly evaluating and adjusting as the person's communication skills improve. Educators must be adaptable, continuously refining the AAC system to make it work best. 

One notable challenge in AAC that can benefit from modern IT technologies revolves around developing communication materials, especially those based on symbols. Symbols are visual representations of words or concepts and are pivotal in bridging communication gaps. However, the source of symbols can be limiting, drawn from libraries with a predefined set that may not align perfectly with individual requirements. Alternatively, educators may manually create symbols, a laborious process demanding graphic design skills.
We need a balanced approach with versatile symbol libraries and user-friendly tools to address this material creation challenge and customize symbols. Collaboration within the AAC community facilitates resource sharing and innovative solutions, enhancing the accessibility and quality of AAC materials.

Foundation models have gained substantial popularity~\cite{zhao2023survey}. These models, powered by artificial intelligence, streamline content creation and offer educators efficient tools to craft personalized learning materials. They adapt to diverse educational needs, making them indispensable for modern education.

This paper explores new ways to use advanced foundation models in AAC. Combining several existing technologies with these models can bring about significant changes in AAC, creating new learning and practical opportunities. However, this approach also brings new challenges related to accessibility, ethics, model biases, and tailoring AAC systems to individual users. By carefully addressing these challenges, we can fully utilize advanced models in AAC and make the world more inclusive for people with communication disabilities.

This paper aims to guide the development of innovative AAC learning methods based on content digitalization and foundation models. To achieve this goal, we first provide an overview of AAC (\autoref{sec:aac_background}) and foundation models (\autoref{sec:foundation}). Subsequently, in \autoref{sec:ambra}, we introduce our approach named AMBRA (\textit{Pervasive and Personalized Augmentative and Alternative Communication based on Federated Learning and Generative AI}), which is designed to address many AAC issues with an open and interoperable approach.
This approach, however, poses new challenges that we analyze in \autoref{sec:challenges}. \autoref{sec:conclusions} concludes the paper with potential future research directions.

AMBRA aims to contribute to creating a more inclusive society, and we hope this paper will inspire enthusiastic research and activity in this field.

\newpage

\markboth{A. Di Paola \MakeLowercase{\textit{et al.}}: Foundation Models in AAC: Opportunities and Challenges}{A. Di Paola \MakeLowercase{\textit{et al.}}: Foundation Models in AAC: Opportunities and Challenges}

\section{A Primer on Augmentative and Alternative Communication (AAC)}\label{sec:aac_background}

AAC refers to a set of tools, strategies, and techniques designed to assist individuals with communication disabilities in expressing themselves effectively~\cite{CUMMINGS202312}. 
AAC is particularly valuable for people who have difficulty with spoken or written language (or both) due to various conditions such as speech and language disorders (including aphasia, apraxia of speech, stuttering, or developmental language disorders), autism spectrum disorders, cerebral palsy, and brain injuries (including traumatic brain injuries or other neurological conditions) that can result in communication impairments AAC can mitigate. 
Enhancing communication, AAC aims to allow individuals with disabilities to convey their thoughts, feelings, needs, and desires, ultimately improving their quality of life and facilitating participation in everyday activities. 

AAC must be highly personalized, with communication solutions tailored to each person's needs and abilities. AAC encompasses diverse tools and techniques based on the individual's specific condition. Simple, low-tech solutions include gesture and sign language, Picture Communication Systems, and communication boards with symbols or pictures. High-tech devices include Speech-Generating Devices (SGDs), Text-to-Speech Software, or tablets for bidirectional communication.
In all cases, \textbf{AAC symbols} facilitate communication for individuals with speech disabilities by visually representing words, phrases, or concepts. These symbols can be custom-designed, selected from extensive picture libraries, follow standardized systems like Blissymbols, PECS (Picture Exchange Communication System), Widgit Symbols, or BoardMaker, or utilize digital symbol sets. Some AAC users also benefit from symbols derived from real photographs, tactile 3D objects, or symbols representing sign language gestures, all tailored to individual needs and preferences. A sequence of symbols can represent complex concepts (e.g., behavioral rules) or \textit{stories}.  

While AI has already been investigated in the context of CAA for speech generation~\cite{konadl2023artificial}, one of the most critical aspects of AAC is indeed the \textbf{content creation}, which is a personalized process carefully aligned with an individual's unique requirements, preferences, and communication skills. The primary aim is to establish a communication mode that is effective and deeply meaningful for the user. The content creation starts with creating the \textit{message}, which is later converted into symbols. The message creation mostly depends on the context and the age, while the selection of symbols is mostly based on the disability level of the given individual.
This selection can encompass standardized symbols, custom-designed symbols, or a combination of both, with the decision on the symbol system being a collaborative effort involving speech-language pathologists, educators, and caregivers. Once a symbol has been associated with a word for a given individual, this \textit{concept} must be preserved every time the individual deals with the same word--at least until the same "concept" can be \textit{generalized} to a new symbol. Typically, educators with expertise in AAC methods must guide and oversee not only the selection and customization process of the symbols but also their use.

\subsection{AAC Challenges}\label{sec:aac_challenges}

AAC is a valuable tool for individuals with communication disabilities but faces at least the following challenges.

\vspace{4pt}\noindent\textbf{Social Barriers.}
Advocacy and awareness efforts must promote the importance of AAC within society. 
Indeed, people who use non-traditional communication methods are often underrepresented in media, literature, and popular culture. So, they may face stereotypes and misconceptions about their cognitive abilities or intelligence. Similarly, peers may be unaware of how to interact with someone who uses AAC. 
Members of the general public, educators, and even healthcare professionals often possess limited knowledge and awareness of AAC systems. The attitudes and comfort levels of communication partners, such as family members, peers, and educators, significantly influence the efficacy of AAC. Negative attitudes or a reluctance to engage with AAC users can be substantial barriers, leading to feelings of isolation and exclusion from conversations or activities. 
Finally, not all public spaces, schools, or workplaces may be well equipped to support AAC systems. For example, boards with AAC symbols are increasingly installed in public places like parks. However, such symbols are fixed and not personalized for different individuals. This underrepresentation can contribute to a lack of understanding and acceptance within society.

\noindent~$\triangleright$~\textit{Possible Approach: AAC can become more extensively used when well supported within society. For this reason, increasing the possibilities of using AAC and creating content more effectively can become a key enabler.}

\vspace{4pt}\noindent\textbf{Technological Barriers.} Communication barriers can arise from technological constraints, compatibility issues, and financial considerations. AAC solutions, particularly high-tech ones, can incur substantial costs, rendering them unaffordable for certain individuals or communities. Additionally, when AAC is used to transform symbols into synthesized speech, selecting an appropriate synthetic voice to convey one's identity can be challenging in noisy or chaotic places.

\noindent~$\triangleright$~\textit{Possible Approach: AAC methods on commercial devices (e.g., smartphones and tablets) rather than custom ones would lower the costs of adopting AAC, broadening the number of potential users but moving the complexity to software~\cite{10.1145/2499149.2499175}.}

\vspace{4pt}\noindent\textbf{Cultural Barriers.}
AAC systems may face limitations in accommodating solutions for individuals coming from different backgrounds, languages, and cultures, which can add complexity to effective communication. For instance, cultural nuances can influence the symbols used for health-related concepts, such as a 'red cross' or a 'red crescent,' varying across different cultures. 
This challenge becomes even more pronounced when extending AAC methods to support immigrant integration. For example, the potential for real-time translation and language support within digital AAC systems can significantly enhance communication for users who require multilingual capabilities.

\noindent~$\triangleright$~\textit{Possible Approach: The global nature of communication and the need for cross-language support also emphasize the importance of digital AAC. By introducing the idea of "concepts," the educator can separate the message from its actual implementation with symbols. This way, the actual symbol can be later customized based on the background and the culture.}

\vspace{4pt}\noindent\textbf{Learning Methods and Educators' Skills.}
Providing training and resources for both AAC users and their communication partners is crucial to ensure optimal system usage.
Indeed, many of these challenges demand a collaborative effort involving professionals, educators, caregivers, and the community. These people often lack the necessary training and support to navigate complex AAC systems effectively.  
Using AAC in educational settings presents challenges, primarily due to inadequate training and limited understanding of best practices among educators in all situations. Furthermore, adapting AAC systems to users' evolving language development poses difficulties, particularly in offering age-appropriate vocabulary. 

\noindent~$\triangleright$~\textit{Possible Approach: Digitizing AAC methods can streamline knowledge-sharing among educators, thereby enhancing training and support mechanisms. Integrating AI models can empower educators to develop personalized materials and expand the applications of AAC methods. This positive feedback loop can boost the adoption of AAC, fostering its utilization across diverse scenarios.}

\vspace{4pt}\noindent\textbf{Flexibility.}
Designing AAC content requires a more adaptable approach. Unlike traditional communication methods, AAC serves a diverse user base with unique communication needs, preferences, and abilities. To ensure its effectiveness, AAC content must be versatile and customizable, enabling the creation of individualized solutions based on the communication styles and proficiency levels. 
The flexibility in AAC content design also accommodates changes over time (e.g., as the person grows), as users' language skills and requirements evolve. This adaptability empowers individuals with communication disabilities by providing them with dynamic and personalized tools. Ultimately, this enhances their quality of life and facilitates active social participation.

\noindent~$\triangleright$~\textit{Possible Approach: There is a clear and increasing demand for the extensive digitalization of AAC methods. The pervasive use of digital devices has naturally aligned with AAC requirements. Given the prevalence of smartphones, tablets, and computers, individuals with communication impairments can greatly benefit from AAC solutions that leverage the capabilities of these devices.}

\vspace{4pt}\noindent\textbf{Data Analysis Methods.}
The advancements in AI methods have opened up new possibilities for analyzing the information used in AAC methods. AAC systems must be more and more adaptive, responsive, and context-aware to improve the user experience and effectiveness of communication for individuals with speech and language challenges. Moreover, personalizing and tailoring AAC solutions to individual needs requires access to customized vocabularies, voices, and communication strategies. So, AAC methods may demand data collection and analysis, which can lead to insights for improving communication interventions and therapies. By capturing these data, researchers and clinicians can gain valuable insights into the effectiveness of AAC solutions and refine them better to meet the needs of individuals with communication disabilities.

\noindent~$\triangleright$~\textit{Possible Approach: Using AI-based methods can support the analysis of the interactions between individuals and AAC devices and suggest content changes. This information can be used as feedback to improve the given model.}

\section{Foundation Models for Content Generation}\label{sec:foundation}

This paper focuses on how foundation models can enhance digital AAC to generate personalized material.
Foundation models are large pre-trained machine learning models that can be classified into language and vision models~\cite{zhao2023survey}. \textbf{Language Models} are examples of foundation models for natural language processing. They can understand and generate text, making them useful for language translation, text summarization, question-answering, sentiment analysis, etc. \textbf{Vision Models} are examples of foundation models for computer vision tasks. These models can analyze and generate images, making them useful for classification, object detection, image generation, etc. This paper focuses on how the models can be used for content creation from textual descriptions.

These models are typically trained on massive datasets (often with hundreds of gigabytes or even terabytes of data) and contain billions of parameters~\cite{shanahan2023talking}. 
After pre-training on this data, these models can be fine-tuned for specific applications with smaller, task-specific datasets. 
This fine-tuning process saves developers and researchers time and resources compared to training models from scratch.

\begin{table*}[t]
\renewcommand{\arraystretch}{1.5}
\centering
\caption{Comparison of foundation models for the creation of AAC material}
\label{tab:llm-compare}
\vspace{-2pt}
\begin{tabular}{@{}C{0.5cm} @{}C{3.5cm} C{6.cm} C{6.5cm}@{}}
\toprule
  &
 \multicolumn{1}{c}{\textbf{Model}} &
 \multicolumn{1}{c}{\textbf{Characteristics}} &
 \multicolumn{1}{c}{\textbf{Use Cases}} \\
\midrule
\multirow{8}{*}{\rotatebox[origin=c]{90}{\parbox[c]{6cm}{\centering Text-to-text generation}}} & Seq2Seq & Encoder-decoder architecture, widely used for various text-to-text tasks & Machine translation, text summarization, dialogue generation, and more \\
& Transformer-based & Self-attention mechanisms, strong performance, scalability & Language modeling, text generation, translation, summarization, question answering, and more \\
& BERT~\cite{devlin2019bert} & Pre-trained model for language understanding, fine-tuned for text generation & Text completion, text generation, text classification, and more \\
& T5 & Designed explicitly for text-to-text tasks, versatile & Text summarization, translation, question answering, and various NLP tasks \\
& RNNs~\cite{recurrent} & Sequential processing, suitable for tasks with temporal dependencies & Text generation, handwriting generation, speech recognition, and more \\
& Hybrid & Combines elements from different architectures for improved performance & Customized for specific tasks, leveraging strengths of multiple approaches \\
& Copy Mechanisms and Attention & Copy mechanisms allow direct copying from input, attention mechanisms focus on relevant information & Summarization, text generation with specific content preservation requirements \\
& Ensemble & Combines predictions from multiple models for enhanced accuracy and robustness & Improving overall performance in text generation tasks \\
\midrule
\multirow{8}{*}{\rotatebox[origin=c]{90}{\parbox[c]{6cm}{\centering Text-to-image generation}}} & cGANs & Utilizes the GAN architecture and conditions image generation on textual descriptions & Generating images from textual descriptions, image-to-image translation, fine-grained image synthesis, and more \\
& VAEs & Employs an encoder-decoder architecture with a continuous latent space & Image generation, data denoising, style transfer, representation learning, and more \\
& Transformer-based & Leverages the transformer architecture and is adapted for text-to-image generation & Generating images from textual descriptions, creative image synthesis based on text prompts, and more \\
& Attribute-Conditioned & Focuses on specific attributes or characteristics mentioned in text to guide image generation & Generating images with desired attributes (e.g., "red car") from textual descriptions and more \\
& Word2Vec+CNN & Combines textual embeddings (Word2Vec) with Convolutional Neural Networks (CNNs) for images & Generating images that correspond to textual descriptions, particularly when semantic embeddings are important \\
& Hierarchical & Employs a two-step generation process, involving scene layout generation and object placement & Coherent scene generation, story illustration, complex image synthesis, and more \\
& Attention & Utilizes attention mechanisms to align text and image features, improving the quality of generated images & Enhancing the alignment between textual descriptions and generated images for better visual quality \\
& Dataset-Specific & Tailored models designed for specific datasets with unique characteristics & Achieving state-of-the-art performance on specific benchmark datasets (e.g., COCO or CLEVR) and more \\
\bottomrule
\end{tabular} 
\end{table*}

Generative AI technologies hold great promise for improving and expanding communication options for individuals with speech and language disabilities~\cite{10.1145/3544548.3581560}.  
In AAC, foundation models can be used to generate content (i.e., creating messages like stories or behavioral rules) and personalized symbols. In the former case, we refer to \textit{text-to-text generation}~\cite{li2021pretrained}, where the model response is a text that represents the AAC message. In the latter case, we refer to \textit{text-to-image generation}, where the model can be used to generate personalized symbols based on the individual background, culture, and abilities.

\subsection{Text-to-Text Generation}

Text-to-text generation is a crucial task in \textit{natural language processing} (NLP) to convert input text (called ``prompt'') into meaningful and contextually relevant output text (called ``response'')~\cite{10.1145/3560815}. This capability can be used in many applications.

\textbf{Text Generation} encompasses the creation of human-like text. These applications contribute to developing AI systems that can communicate with humans more effectively. \textit{This can be one of the key technologies in digital AAC to create new material that is personalized to the user's needs and skills.}
Text-to-text generation also finds relevance in \textbf{Machine Translation}, i.e., translating text from one language to another, facilitating cross-lingual communication, and making information accessible to individuals from different linguistic backgrounds~\cite{zhu2023multilingual}. \textit{This can support multi-language and multi-cultural uses of the AAC methods, where the educators can focus on the contents rather than on the language.}

Other applications are \textbf{Text Summarization} (i.e., the creation of concise and coherent summaries for lengthy pieces of text), \textbf{Question-Answering} (i.e., the generation of accurate and contextually relevant answers), \textbf{Paraphrasing} (i.e. the production of alternatives of a given text), \textbf{Code Generation} (i.e., the generation of code snippets or scripts), and \textbf{Data-to-Text applications} (i.e., the conversion of structured data into human-readable descriptions or reports). 

\vspace{4pt}\noindent
\textbf{Models.} 
Text-to-text generation models are typically based on neural network architectures, such as \textit{transformers}, \textit{recurrent neural networks} (RNNs), or \textit{convolutional neural networks} (CNNs). These models can be trained in a supervised manner with pairs of input and target texts. They can be fine-tuned to adapt the model to a particular text-generation task. \textit{In our approach, we aim indeed to generate an open and domain-specific model for AAC content generation~\cite{touvron2023llama}.}
We discuss the following models: 

\begin{itemize}[leftmargin=8pt]

\item \textbf{Seq2Seq} models (or encoder-decoder models) consist of two main components: an encoder that encodes the input text into a fixed-length vector representation and a decoder that generates the output text based on this representation. 

\item \textbf{Transformer-based} models, such as GPT (Generative Pre-trained Transformers), rely on self-attention mechanisms and positional encodings to process input sequences and generate output sequences. GPT-3, for example, can generate human-like and coherent text from a given prompt and has demonstrated strong performance across multiple domains.

\item \textbf{BERT} (Bidirectional Encoder Representations from Transformers) is a pre-trained transformer-based model that can be used for text-to-text generation by conditioning the model following a given input. 

\item \textbf{T5} (Text-to-Text Transfer Transformer) is a model architecture designed explicitly for text-to-text tasks. It frames all NLP tasks as a text-to-text problem, where both the input and output are treated as text.

\item \textbf{RNNs} (Recurrent Neural Networks) are a class of neural networks that have been used in tasks where sequential dependencies are essential, like text completion~\cite{recurrent}.

\item \textbf{Hybrid} models combine elements from different approaches (e.g., a transformer-based encoder and an RNN-based decoder) to leverage their respective strengths. 

\item \textbf{Copy mechanisms} allow models to copy words or phrases directly from the input text to the output. \textbf{Attention mechanisms} help the model focus on relevant input parts when generating the output.

\item \textbf{Ensemble} models combine the predictions of multiple generation models to produce more robust and accurate results.
\end{itemize}
Each of these text-to-text generation methods has its advantages and is suitable for different use cases. \autoref{tab:llm-compare} compares the different models for text-to-text generation.
The choice of method depends on the specific task requirements, available data, and desired performance metrics. \textit{AAC requires context-aware text generation, especially for the generation of the contents to be later translated into symbols. For this reason, transformer-based and BERT models seem more appropriate.}

\subsection{Text-to-Image Generation}

Text-to-image generation can create images from textual descriptions or captions. While it has the power to create any kind of visual art (from icons to realistic photos), it heavily relies on the expressiveness of the given textual content.

\textbf{Image Generation from Descriptions}. is a remarkable application to generate vividly articulate scenes, objects, or abstract concepts from verbal expressions. \textit{In the context of digital AAC, it can be used to generate personalized symbols for users based on textual content related to symbol description, context, and user information.} Text-to-image generation can also improve \textbf{Accessibility}. Indeed, it can help automatically translate textual descriptions into visual content for individuals with visual impairments. \textit{In a long-term scenario, this method can be used to automatically translate already-existing contents (e.g., parts of websites or public signs) into personalized sequences of symbols.}
\textbf{Content Creation} and \textbf{Visual Storytelling}  are other important applications that can generate a sequence of images for stories, messages, or, in general, any textual story or narrative. \textit{This is a key technology to create new material for AAC users, starting from educators' requests or existing books.}

Other applications include \textbf{Concept Visualization} (i.e., the creation of concrete visual representations from abstract ideas) and 
\textbf{Data Augmentation} (i.e., the possibility to enrich datasets with supplementary images from annotations or labels).

\begin{figure}[!t]
     \centering
     \begin{subfigure}[b]{0.3\columnwidth}
         \centering
         \includegraphics[width=\textwidth]{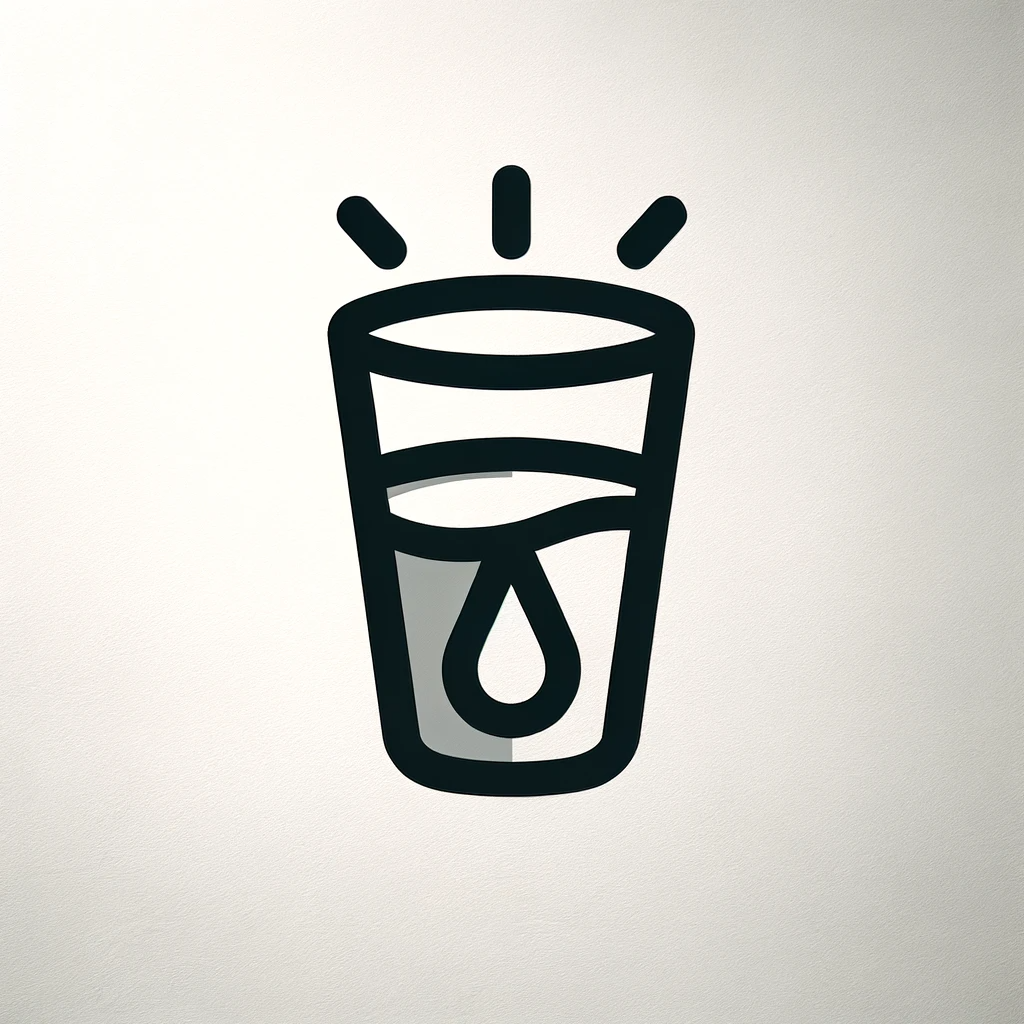}
         \caption{}
         \label{fig:symb1}
     \end{subfigure}
     \hfill
     \begin{subfigure}[b]{0.3\columnwidth}
         \centering
         \includegraphics[width=\textwidth]{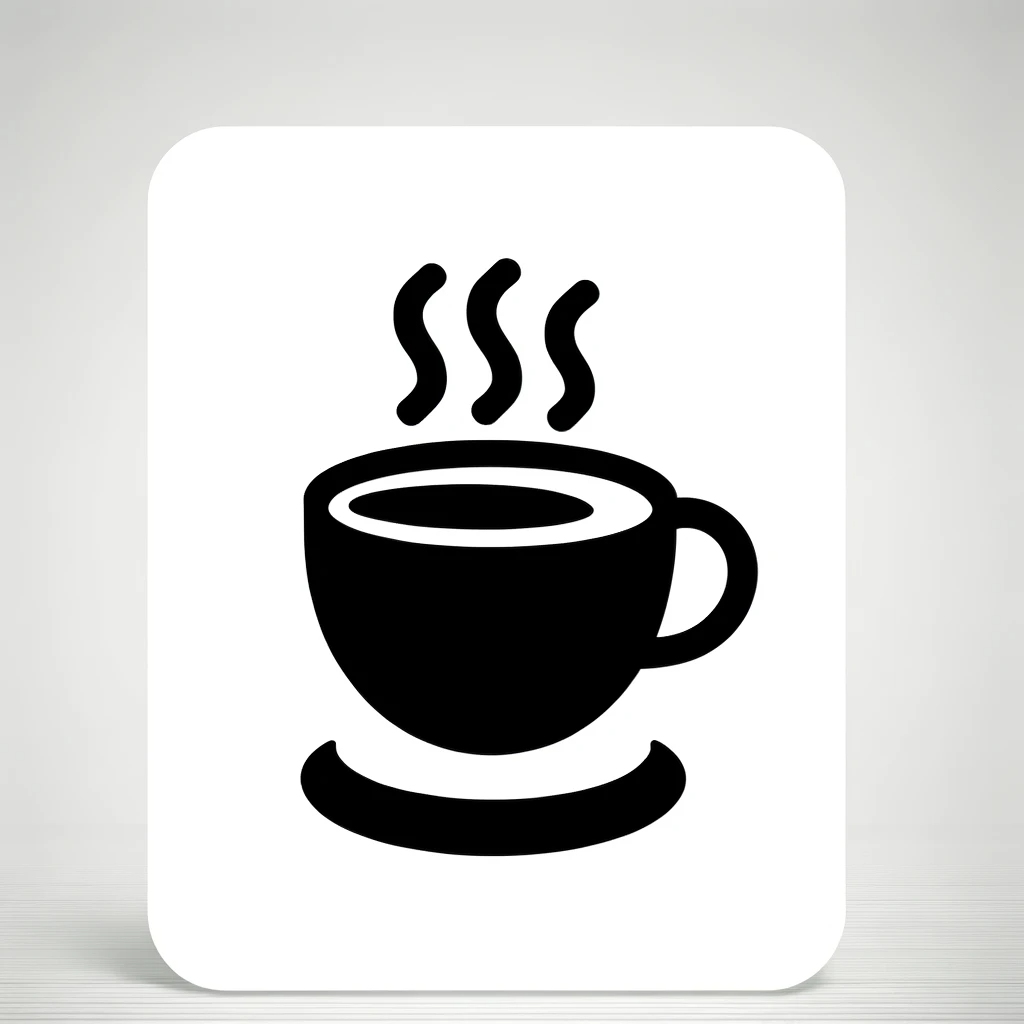}
         \caption{}
         \label{fig:symb2}
     \end{subfigure}
     \hfill
     \begin{subfigure}[b]{0.3\columnwidth}
         \centering
         \includegraphics[width=\textwidth]{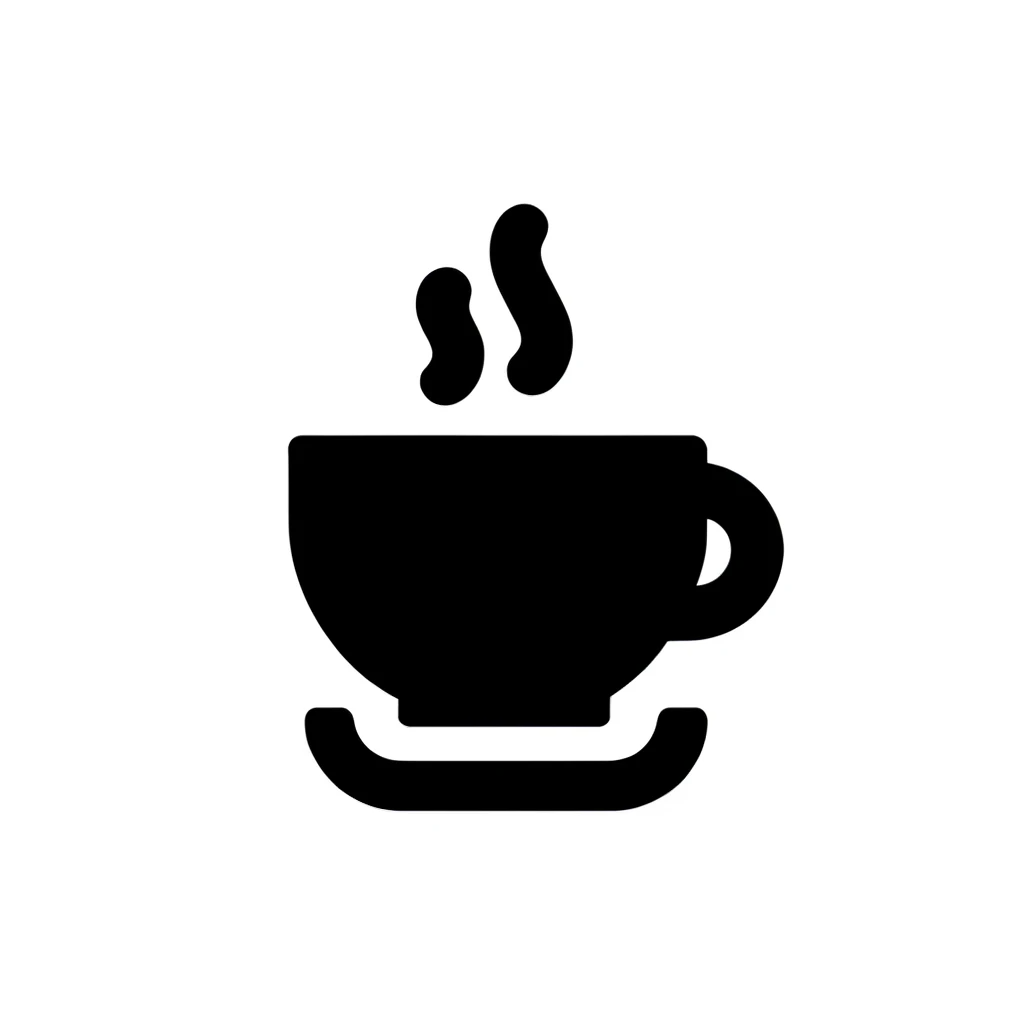}
         \caption{}
         \label{fig:symb3}
     \end{subfigure}
     \centering
     \begin{subfigure}[b]{0.3\columnwidth}
         \centering
         \includegraphics[width=\textwidth]{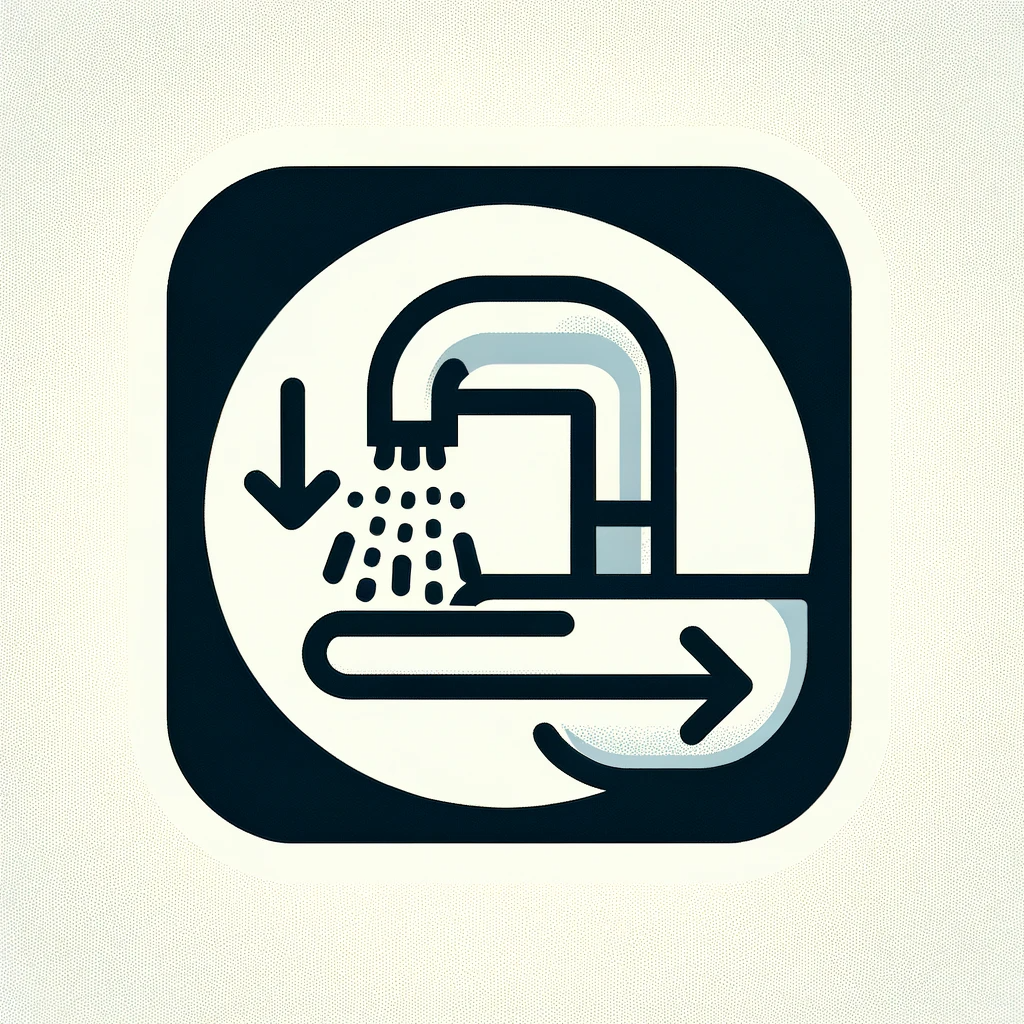}
         \caption{}
         \label{fig:symb4}
     \end{subfigure}
     \hfill
     \begin{subfigure}[b]{0.3\columnwidth}
         \centering
         \includegraphics[width=\textwidth]{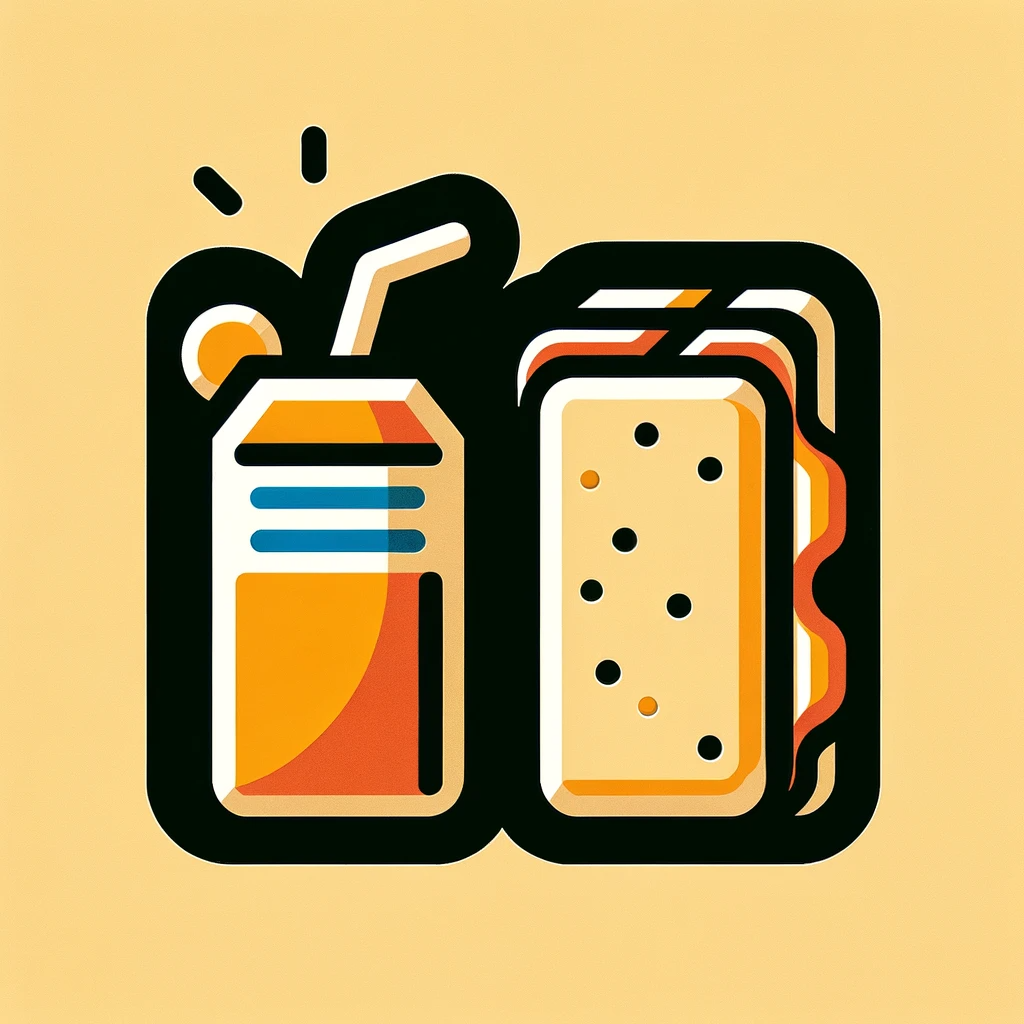}
         \caption{}
         \label{fig:symb5}
     \end{subfigure}
     \hfill
     \begin{subfigure}[b]{0.3\columnwidth}
         \centering
         \includegraphics[width=\textwidth]{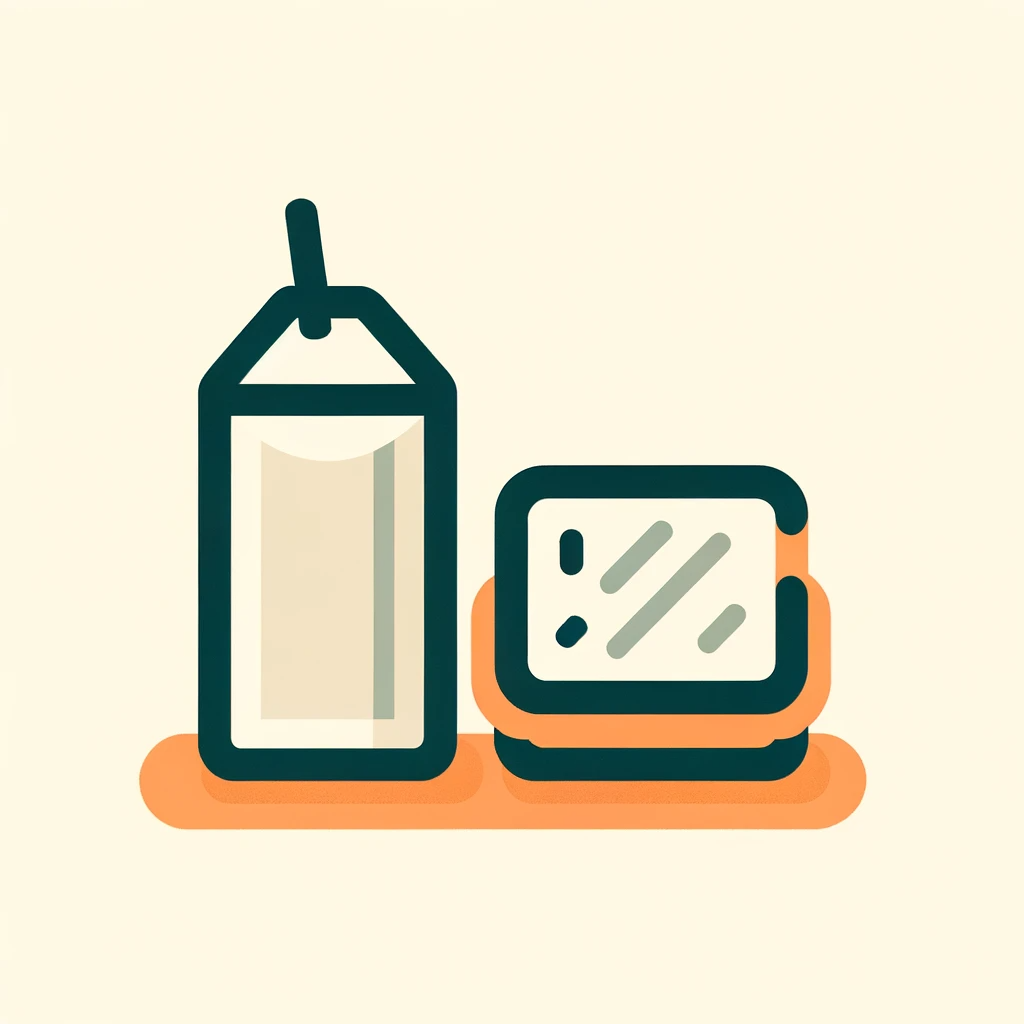}
         \caption{}
         \label{fig:symb6}
     \end{subfigure}
\caption{Examples of AAC symbols created with DALL·E~\cite{dalle}, a text-to-image, transformed-based model developed by OpenAI: a) glass of water; b) cup of coffee; c) simple cup of coffee; d) ``open the sink'' action; e) snack; f) simple snack.}\label{fig:dalle}
\end{figure}

\vspace{4pt}\noindent
\textbf{Models.}
Text-to-image generation models typically employ deep learning techniques, such as generative adversarial networks (GANs), variational autoencoders (VAEs), or transformers to translate textual descriptions into the corresponding images. We discuss the following models: 

\begin{itemize}[leftmargin=8pt]
\item \textbf{cGANs} (Conditional Generative Adversarial Networks) involve two neural networks: a generator and a discriminator. The generator takes a text description as input and generates an image, while the discriminator evaluates the generated image's realism. The two networks are trained adversarially, with the generator aiming to produce images that can fool the discriminator. 

\item \textbf{VAEs} (Variational Autoencoders) are combined with a conditional setup. In this approach, the encoder network maps textual descriptions to a latent space, and the decoder generates images from points in the latent space. VAEs provide a probabilistic framework for generating diverse images from the same text input.

\item \textbf{Transformer-based} models can be conditioned on text inputs to generate images. DALL·E, for example, is a GPT-based model that generates images from textual descriptions.

\item The \textbf{Attribute-Conditioned} models use attribute classifiers to identify and incorporate relevant attributes into the image generation process. For instance, if a text describes a ``red car,'' the model focuses on generating an image where the car is actually red.

\item The \textbf{Word2Vec+CNN} fusion approach combines Word2Vec embeddings of textual descriptions with Convolutional Neural Networks (CNNs) for image generation. Textual embeddings are used to condition the CNN, allowing the model to generate images that correspond to the given text.

\item The \textbf{Hierarchical} models use a hierarchical structure to generate images from text. They first generate a scene layout or structure based on the text and then populate the scene with objects and details. This two-step process can lead to more coherent and realistic image generation.

\item \textbf{Attention} mechanisms are employed to align the generated image elements with the corresponding words in the text. Attention helps the model understand which parts of the text influence the generation of specific image details.

\item In \textbf{Dataset-Specific} approaches, the generation models are designed with specific datasets in mind, such as COCO or CLEVR. These models are tailored to the characteristics and requirements of the particular dataset, which can lead to better performance on it.
\end{itemize}
State-of-the-art models in text-to-image generation have made significant progress in generating high-quality and coherent images that match the provided text descriptions. The choice of approach often depends on the specific requirements of the text-to-image generation task and the available resources.

\autoref{fig:dalle} shows a set of AAC symbols generated with DALL·E~\cite{dalle}, a 16-billion parameter, transformed-based model developed by OpenAI. We generated two versions for two symbols (``cup of coffee'' and ``snack''), requesting the second to be a simplified version of the first. While the symbols are close enough to traditional AAC symbols (see \autoref{fig:aac_example}), their graphics are still too complex for most AAC users. 
This experiment underscores the importance of having a domain-specific model for generating AAC symbols.

\begin{figure*}[!t]
\centering
\includegraphics[height=4cm]{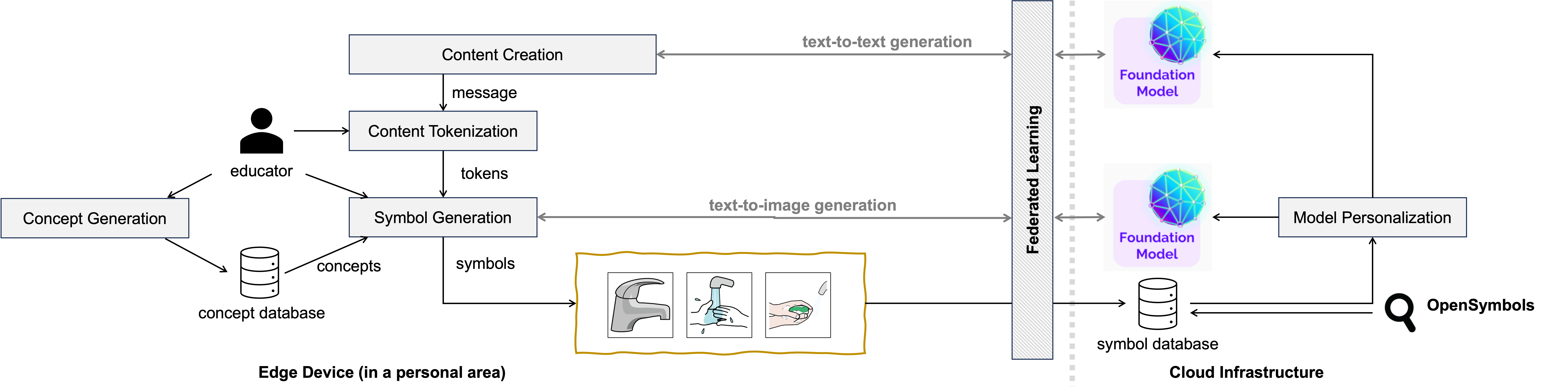}
\caption{Overview of our AMBRA open platform}\label{fig:ambra}
\end{figure*}

\section{AMBRA: Using Foundation Models for Pervasive and Personalized AAC}\label{sec:ambra}

This section presents our concept for digital AAC using foundation models to enhance content creation and personalization. Our approach, called AMBRA (\textit{Pervasive and Personalized Augmentative and Alternative Communication based on Federated Learning and Generative AI}), is shown in \autoref{fig:ambra}. AMBRA is an open platform, i.e., a modular and extensible framework, that can combine several technologies to address the challenges described in \autoref{sec:aac_challenges}. The key idea is to create a fine-tuned, domain-specific model for AAC (to support problem-specific creation) that can serve several individual-specific requests (to support personalized creation)~\cite{wei2022finetuned}. Combining a cloud-based infrastructure with common edge devices (e.g., desktops, laptops, tablets, smartphones), AMBRA can simplify content creation, support knowledge sharing among educators, and pave the way to new AAC learning methods, either digital or paper-based. The resulting infrastructure can extend the use of AAC within society, opening new opportunities but also challenges (see \autoref{sec:challenges}). In the following, we describe the major AMBRA components, how they can interact with each other, and the requirements for its successful deployment. 

AMBRA focuses on personalized content creation and visualization for AAC users. This requires each educator to have a \textit{profile} containing information about the users he/she is working with. The educator can import existing content (shared by other educators, already created for other users, or publicly available as a ``content package'') for each user. Note that, since symbols are personalized for the given user, the content will include only information about the AAC message (e.g., social story or behavioral rule). The creation of new material is enhanced with foundation models and performed in a multi-platform app that is developed in Dart and so based on the Flutter framework. This allows us to keep a single, open code base and deploy the application on all common devices. Such an application can be later extended with features based on the specific needs of the educators or provide requests to the most common cloud infrastructures (e.g., Amazon AWS or Microsoft Azure) with stable APIs.

In \textbf{Content Creation}, educators can manually generate AAC ``content'' (e.g., stories, rules, agendas, etc.) or use text-to-text generation models to automatically generate new material by combining the previous experience on the same individual or similar topics. The \textit{content} is a text message that is later tokenized to be converted into symbols (\textbf{Content Tokenization}). AMBRA proposes alternative token lists based on the symbols currently available in the application library. The educator can select one of these solutions and provide further customization by splitting combined concepts or combining single words into more complex concepts. Once the message tokenization is acceptable for the educator, the tokens are converted into the corresponding symbols (\textbf{Symbol Generation}) for the given user. Symbols are first selected based on the available concepts for the given individual. If the educator had a previous experience with a specific concept, the association is preserved and reproduced for all the new instances. Otherwise, the symbols can be selected from a pre-defined library, added by the educators through camera or image uploading, or generated with text-to-image generation models. This information is added to the user's profile and eventually used as an additional (re-)training set for the foundation model. To preserve the idea of having a fully open solution, we aim to include symbols for open libraries, like OpenSymbols. 
To create meaningful AI prompts, the information of the token is combined with \textit{context} information (provided by the educator or extracted from the message analysis) and the other concepts understood by the same user. Additional context information can be obtained with the interaction with the environment. For example, geotagging information can be used to determine the users' location even though this opens serious privacy issues that must be adequately addressed at the legal and technological level (i.e., with a communication protocol that adheres to strict regulations). Our customization approach is based on \textit{few-shot learning} and allows us to generate new images by extracting features from images that the target individual is known to understand. This greatly increases the probability that the new symbol is correctly interpreted. If not, the educator can provide feedback that is later used to educate the model.

To enforce flexibility and customization, AMBRA can leverage federated learning to create personalized content for the individual, creating a local sub-model for each of them (\textbf{Model Personalization}). These decentralized models are deployed on the educator's computing resources and/or the terminal devices of the individual. However, the information is also shared with the central model (on the cloud) to allow for knowledge sharing among educators. 
Indeed, if many individuals understand the same concept similarly, this information is centralized and becomes part of the domain-specific (and not the individual-specific) model.

\section{AMBRA Opportunities and Challenges}\label{sec:challenges}

Our AMBRA platform opens new opportunities and challenges for AAC. In the following, we discuss the major ones, identifying potential solutions for each of them.

\vspace{4pt}\noindent\textbf{Creation of highly personalized material.} AMBRA is designed to facilitate the creation of highly personalized materials. It accomplishes this by employing a unique approach that separates the message from the symbols through the use of what can be termed ``concepts.'' This decoupling of the message and the symbols provides users with unparalleled flexibility and adaptability when it comes to crafting communication materials tailored to their specific needs. Indeed, the symbols are provided only when the message is ``deployed'' for the given user, selecting the ones that are more appropriate in the specific case. In this way, different users can visualize the message differently based on their capabilities and the educator's experience.
Additionally, AMBRA leverages foundation models to fill in the gaps when the desired symbols are not in the existing symbol repository. This dynamic integration of foundation models ensures that the pre-existing symbol sets do not limit users. Instead, they can access a vast and diverse pool of symbols, further enhancing their ability to communicate and express themselves.

\textit{Challenges and Possible Solutions:} Creating personalized symbols requires driving text-to-image generation models toward the creation of simplified visual arts with specific characteristics. Requests must be generally enriched with precise information extracted from context and educators' hints.
While fine-tuning the models (e.g., with reinforcement learning techniques) can improve image generation quality and performance, they require big data sets that can be achieved only with open symbols and knowledge sharing. Few-shot learning~\cite{brown_nips20} can better drive the generation process when some user's information is already available.
Additional symbols gathered by the educators and caregivers can be used to support retraining of these models but open important privacy issues. Federated learning can help derive knowledge by processing the information locally.

\vspace{4pt}\noindent\textbf{Virtuous circle of open solutions.} AMBRA is an open platform and aims to be built on top of free solutions (i.e., models, software, and libraries). 
While we recognize the effort for crafting proper symbols, this burden can be alleviated only through collaborative efforts and the global sharing of materials among educators and professionals. 
We hope that creating such an open environment can boost the creation of more open symbol libraries (like OpenSymbols. As symbol libraries become progressively richer and more diverse, they become even more beneficial to educators and users, further enhancing the adoption of AAC as an effective means of communication.
By embracing an open environment where educators worldwide can freely exchange and share resources, we aspire to initiate a transformative shift in the AAC landscape so that it can be used extensively in society at low cost. 

\textit{Challenges and Possible Solutions:} Sharing free material is challenging for educators who invested time and energy to create it. They can be further motivated when they operate within an open community. In this context, the standardization of symbols, concepts, and materials is essential to increase reusability. Also, the interfaces among the different components require new standards to help interoperability, including the possibility of upgrading components and models when new solutions become available.

\vspace{4pt}\noindent\textbf{Use of AAC methods in real life.}
Open and inexpensive solutions for creating AAC material can help democratize this technology in society. Such solutions not only make AAC more accessible but also significantly reduce the learning curve for individuals and caregivers seeking to harness these methods for effective communication. Shops, public places, and various community spaces can readily adopt these communication methods to foster inclusivity. Unlike traditional pre-designed AAC materials, these open solutions offer the unique advantage of generating communication content on the fly, catering to each user's specific needs and preferences. 
Modern technologies such as QR codes serve as a bridge between physical spaces and digital content, enabling seamless communication between individuals with diverse communication needs and their surroundings. QR codes can be strategically placed in public places, on product labels, or within businesses, allowing users to scan them with their devices. This action triggers the instantaneous loading of tailored AAC content, ensuring that individuals can effectively communicate their preferences, questions, or needs.
By harnessing geolocation data, the user's device can automatically derive context-aware information, offering relevant AAC content based on the surroundings. For instance, when navigating a shopping mall, the user's device can provide context-specific symbols or phrases, aiding in interactions with store personnel or fellow shoppers.

\textit{Challenges and Possible Solutions:} While inclusiveness is an important motivation, many commercial, public, and community spaces are generally motivated by economic returns. Making the AMBRA technology pervasive at a low cost can motivate more people to invest in these methods. A widespread adoption of AAC methods can create new opportunities for the job market. For example, shopping malls can share employees that keep AAC signs and materials updated for the customers. Privacy issues in public places can be addressed by adopting federated learning. New regulations and policies for sharing personal information must be developed when using it in AI-enhanced systems, especially to make them compliant with the most recent General Data Protection Regulation (GDPR).

\vspace{4pt}\noindent\textbf{New profiles of educators.}
Currently, educators are requested to learn the best practices and tools with limited knowledge sharing and resources. Not all organizations can access expensive tools and devices, which can significantly constrain the effectiveness of AAC design methods. By providing educators and caregivers with open methods and resources, we can fundamentally alter the educational landscape. 
These professionals can channel their energy and expertise toward what truly matters--the content itself. They are not required anymore to spend hours designing new symbols, but they can take suggestions from AI-based systems. This paradigmatic shift can lead to the creation of new educator profiles that combine pedagogical, psychological, and technological skills. These new educators must be able to steer these new technologies to create the proper content, using them as a tool and being aware of potential and limitations.

\textit{Challenges and Possible Solutions:} Educational profiles for social and technological matters are usually independent of each other. AMBRA advocates the creation of interdisciplinary environments where, for example, pedagogical and linguistic skills are combined with ICT skills. New courses or masters can be developed to show how to combine such skills. Experienced educators should offer extensive training to provide guidelines on using these technologies.

\vspace{4pt}\noindent\textbf{Use of AAC methods in other contexts.}
The design of effective digital AAC methods can extend their adoption in several contexts. For example, digital AAC methods allow deaf individuals with linguistic impairments to communicate effectively. Traditional sign language may not always be accessible or sufficient, especially when interacting with individuals who do not understand it. Young immigrants can leverage more universal symbols to start communication with peers and teachers, facilitating the initial stages of language acquisition. In both these scenarios, digital AAC methods support effective communication and integration. They leverage technology to break down barriers and create pathways for individuals to express themselves and connect with others. Furthermore, these applications showcase the versatility of digital AAC, extending its potential to impact various domains.

\textit{Challenges and Possible Solutions:} The creation of multimodal, multicultural, and multilingual material is challenging. This requires additional context-aware information, resulting in larger retraining that demands high resources and large datasets. On the one hand, it creates new possibilities for educators to develop innovative learning methods based on symbols and games. On the other hand, it requires extensive use in society to be impactful.

\vspace{4pt}

In conclusion, AMBRA's idea extends beyond its technical architecture; it embodies a vision for a more inclusive future where the boundaries of communication are pushed ever further thanks to the vicious circle created by the combination of all these solutions, unlocking the potential for individuals with diverse communication needs to express themselves with clarity and dignity in any context and without barriers.

\section{Conclusions and Future Directions}\label{sec:conclusions}

This paper discussed the potential use of foundation models in augmentative and alternative communication to boost content creation and allow the pervasive use of this technology in society. We presented AMBRA, our open platform to support educators. AMBRA can seamlessly integrate several technologies (e.g., generative AI, federated learning, positioning, secure communication, etc.), opening new possibilities for AAC learning methods. This approach can lead to a rethink of the educator's role to be more open to worldwide collaboration in material and knowledge sharing.
So, this paper discusses the opportunities and the challenges for this approach. It aims to be an open approach that calls for collaboration rather than competition. For this reason, this paper will be periodically revised by including any feedback that we receive on this idea.

\section*{Acknowledgements}

The authors would like to thank Azalia Mirhoseini (Stanford University), David Atienza (EPFL), Roberta Piscitelli (Amazon AWS), Alberto Sangiovanni-Vincentelli (University of California at Berkeley), Franca Garzotto (Politecnico di Milano), and Manuel Roveri (Politecnico di Milano) for the valuable and insightful discussions about this work.

\printbibliography

\end{document}